\documentclass[seceq]{ptptex}
\makeatletter
\def\vereq#1#2{\lower3pt\vbox{\baselineskip1.5pt \lineskip1.5pt
\ialign{$\m@th#1\hfill##\hfil$\crcr#2\crcr\sim\crcr}}}
\makeatother
\def\lsim{\mathrel{\vcenter{\hbox{$<$}\nointerlineskip\hbox{$\sim$}}}}
\title{
 Fermion Mass Hierarchy and Supersymmetry Breaking 
in $6D$ $SO(10)$ GUT on Orbifold
}
\author{
Naoyuki {\sc Haba$^{1,2}$}\footnote{haba@eken.phys.nagoya-u.ac.jp}
Yasuhiro {\sc Shimizu$^2$}\footnote{shimizu@eken.phys.nagoya-u.ac.jp}
}
\inst{
 $^1$Faculty of Engineering, Mie University, Tsu, Mie,
  514-8507, Japan\\
 $^2$Department of Physics, Nagoya University, Nagoya, 
  464-8602, Japan
}

\abst{
We suggest simple models which produce the suitable fermion
 mass hierarchies and flavor mixing angles based on the 
 6 dimensional $N=1$ supersymmetric $SO(10)$ grand unified theory 
 compactified on a $T^2/(Z_2 \times Z_2')$ orbifold.
We introduce extra vector-like heavy fields in the extra 
 dimensions, and  
 the suitable fermion mass hierarchies and flavor mixings 
 are generated by integrating out 
 these heavy fields.
We consider gaugino mediation and gauge mediation 
 supersymmetry breaking mechanisms and their flavor 
 structures. 
The experimental constraints of small 
 flavor changing neutral currents 
 suggest where to locate 
 the supersymmetry breaking brane 
 in the gaugino mediation mechanism. 
On the other hand, 
 the SUSY breaking 
 masses are highly 
 degenerated in the gauge mediation scenario, 
 where the flavor changing neutral currents 
 are naturally suppressed as in the ordinal 
 four dimensional gauge mediation models. 
}

\begin{document}

\maketitle

\section{Introduction}
\label{sec:intro}

Grand unified theories (GUTs) are very attractive models in which the three
 gauge groups are unified at a high energy scale. 
However, one of the most serious problems to construct a model of GUTs 
 is how to realize the mass splitting  between the triplet and 
 the doublet Higgs particles  in the Higgs sector. 
This problem is so-called  triplet-doublet (TD) splitting problem. 
A new idea for solving the TD splitting problem 
 has been suggested in higher dimensional 
 GUTs where the extra dimensional 
 coordinates are compactified on 
 orbifolds.\cite{5d}\tocite{so(10)others}
In these scenarios, 
 Higgs and gauge fields are propagating in
 extra dimensions, and  
 the orbifolding realizes the gauge group
 reduction and 
 the TD splitting since  
 the doublet (triplet) Higgs fields have (not)
 Kaluza-Klein zero-modes. 
A lot of attempts and progresses have been done in 
 the extra dimensional GUTs on orbifolds.\cite{6d}\tocite{Shafi:2002ii}
Especially, the reduction of $SO(10)$ gauge symmetry 
 and the TD splitting solution are 
 first considered in 6D models in Refs.~\citen{ABC}, \citen{HNOS}.

As for producing fermion mass hierarchies,  
 several trials have been done 
 in the extra dimensional GUTs on 
 orbifolds.\cite{HSSU}\cite{HKS}\cite{flavor}\cite{Hebecker:2002re}
\cite{comple}\cite{SO(10)SU(6)gauge}\cite{Hebecker:2002xw}
\cite{Shafi:2002ii}
The model in Ref.~\citen{HKS} can induce the natural
 fermion mass hierarchies and flavor mixings 
 based on a 6D $N = 1$ SUSY ((1,0)-SUSY) 
 $SO(10)$ GUT where
 the 5th and 6th dimensional coordinates are 
 compactified on a $T^2/Z_2$ orbifold.
In this scenario, we introduce extra vector-like 
 generations, $2 \times ( \psi_{\mathbf{16}_4}+ 
 \psi_{\overline{\mathbf{16}_4}})$ 
 and $( \psi_{\mathbf{16}_5}+ \psi_{\overline{\mathbf{16}_5}})$, 
 which propagate 6 and 5 dimensions, respectively. 
Assuming that 
 4th (5th) generation vector-like fields only couple to 
 the 1st (2nd) generation chiral fields, 
 the suitable fermion mass hierarchies and flavor mixings 
 are generated by integrating out
 these vector-like heavy fields. 
The mixing angles between the chiral fields and 
 extra generations have been determined by
 the volume suppression factors. 
The extension of this model has 
 been considered in Ref.~\citen{HKS2}
where the values of $m_e$, $m_d$, $V_{us}$, and $V_{e3}$ 
 have been improved by extending the vector-like extra generations
 and their configurations in the extra dimensions. 
However, there is a difficulty in this scenario. 
That is the lack of 5D fixed lines, which can not guarantee 
 $( \psi_{\mathbf{16}_5}+ \psi_{\overline{\mathbf{16}_5}})$ existing 
 only in 5 dimensions, not spreading in 6 dimensions\footnote{We 
 would like to thank T. Kugo for pointing out this 
 problem.}.

In this paper we will modify previous papers\cite{HKS}\cite{HKS2} 
 by using the orbifold, $T_2/(Z_2 \times Z_2')$, and 
 also consider the SUSY breaking mechanism. 
We will consider 
 the 6D $N=1$ $SO(10)$ GUT with vector-like 
 matter contents on $T_2/(Z_2 \times Z_2')$.  
As will be shown bellow, this modification 
 makes no changes for the zero mode matter 
 fields in Refs.~\citen{HKS}\citen{HKS2}, so that 
 the mechanism of creating the fermion mass hierarchies 
 in the previous papers are useful. 
The gauge symmetry reduction 
 and the TD splitting are also the same as those in the 
 $T_2/Z_2$ orbifold. 
The gauge and Higgs fields live in 6 dimensions
 and the orbifolding and boundary conditions make the $SO(10)$ 
 gauge group be broken to 
 $SU(3)_C\times SU(2)_L \times U(1)_Y \times U(1)_X$ and
 realize the TD splitting.

As for the SUSY breaking mechanisms, 
 we will consider 
 the gaugino and the gauge mediation scenarios. 
In the gaugino mediation scenario, 
 the vector-like matter fields in extra dimensions
 can directly couple to the SUSY breaking fields,
 which induces non-universal contributions to
 SUSY breaking masses for the light matter fields.
These non-universal SUSY breaking masses can give rise to
 too large flavor changing neutral currents (FCNCs).  
Thus, the location of the SUSY breaking brane 
 should be determined in order to avoid the large FCNC 
 phenomenological problems in the gaugino mediation 
 scenario. 
On the other hand, 
 the SUSY breaking 
 masses for the light matter fields are highly 
 degenerated in the gauge mediation scenario, 
 where the FCNCs 
 are naturally suppressed as in the ordinal 
 4D gauge mediation models.

\section{Fermion mass hierarchies and flavor mixings}

We consider the 6D $N=1$ SUSY 
 $SO(10)$ GUT, whose  
 extra dimensional coordinates are compactified 
 on a $T^2/(Z_2 \times Z_2')$ 
 orbifold.\cite{HNOS}\cite{SO(10)SU(6)gauge}
The structure of extra 2D spaces are 
 characterized by reflection $P$ $(Z_2)$, $P'$ $(Z_2')$, 
 and translations $T_i$ ($i = 1, 2$). 
Under the reflection $P$ and $P'$, 
 $(z, \bar{z})$ is transformed into 
 $(-z, \bar{z})$ and $(z, -\bar{z})$, respectively. 
Where $z \equiv (x_5+ix_6)/2$ and $\bar{z} \equiv (x_5-ix_6)/2$
 with the physical space of 
 $0 \leq x_5, x_6 < \pi R$. 
Under the translation 
 $T_1$ and $T_2$, $(z, \bar{z})$ are transformed into 
 $(z + 2\pi R_z, \bar{z})$ and 
 $(z, \bar{z} + 2\pi R_{\bar{z}})$, respectively,
 where $R_z\equiv (1+i)R/2$ and $R_{\bar{z}}\equiv (1-i)R/2$.  
The physical space can be taken as $0 \leq z < \pi R_z$ and 
 $0 \leq \bar{z} \leq \pi R_{\bar{z}}$. 
Thus, the $T_2/(Z_2 \times Z_2')$ orbifold is just 
 the same as the $S_1/Z_2 \otimes S_1/Z_2'$ orbifold
 of a regular square.  
There are four fixed points at 
 $(0,0)$, $(\pi R_z,0)$, $(0, \pi R_{\bar{z}})$ and 
 $(\pi R_z, \pi R_{\bar{z}})$, 
 and two fixed lines on $z=0$ and $\bar{z}=0$ 
 on the orbifold. 
The bulk fields are decomposed by $P$, $P'$, and $T_i$. 
{}For examples, 
 a 6D bulk scalar field $\Phi(x^\mu, z,\bar{z})$ is 
 decomposed into
\begin{eqnarray}
\Phi_{(\pm \pm )(\pm \pm)}(x^{\mu},z,\bar{z}) &\equiv& 
 \frac{1}{\pi R_c}
   \phi_{(\pm \pm)_{z}(\pm \pm)_{\bar{z}}}(x^\mu)
    \varphi_{(\pm \pm)_{z}}(z)
    \varphi_{(\pm \pm)_{\bar{z}}}(\bar{z}), 
\end{eqnarray}
according to the eigenvalues 
 of $(P,T_1)(P',T_2)\; (=(P,T_1)_{z}\otimes(P',T_2)_{\bar{z}})$.  
Where $R_c \equiv |R_z| (=|R_{\bar{z}}|)$. 
Notice that only $\Phi_{(+\pm )(+\pm )}$ can have massless 
 zero-modes 
 and survives in the low energy.

We consider 
 the gauge multiplet and two ${\mathbf 10}$ representation 
 Higgs multiplets propagate in the 6D bulk,
 which are denoted as ${\bf H_{10}}$ and ${\bf H'_{10}}$,
 and the ordinal 
 three-generation matter multiplets 
(${\mathbf{16}}_i, \;i=1,2,3$) are
 localized on the 4D brane, $(0,0)$. 
We adopted the translations 
 as $T_{51} = \sigma_2\otimes I_5$ and 
 $T_{5'1'} = \sigma_2\otimes diag.(1,1,1,-1,-1)$, 
 which commute with the generators
 of the Georgi-Glashow $SU(5)\times U(1)_X$\cite{GG} and 
 the flipped $SU(5)'\times U(1)'_X$\cite{Fl} groups, 
 respectively.\cite{ABC}\cite{HNOS}
Then, translations ($T_i$) make the $SO(10)$ 
 gauge group be broken to 
 $SU(3)_C\times SU(2)_L \times U(1)_Y \times U(1)_X$ and
 realize the TD splitting
 since the doublet (triplet) Higgs fields have (not)
 Kaluza-Klein zero-modes.

The zero mode of the 
 6D bulk matter field, 
 ${\psi_{\bf 16}}_{(+\pm )(+\pm)}$,  
 is classified into four types as 
\begin{eqnarray}
 {\psi_{\bf 16}}_{(++)(++)}&~~~&\mbox{(zero mode)~}=Q, \nonumber\\
 {\psi_{\bf 16}}_{(++)(+-)}&~~~&\mbox{(zero modes)}=\overline{U}
,\overline{E}, \nonumber\\
 {\psi_{\bf 16}}_{(+-)(++)}&~~~&\mbox{(zero modes)}=\overline{D},
\overline{N}, \nonumber  \\
 {\psi_{\bf 16}}_{(+-)(+-)}&~~~&\mbox{(zero mode)~}=L. 
\label{6d}
\end{eqnarray}
Similarly, 
 the zero mode of the 5D bulk field, 
 which is propagating on the fixed line $\bar{z} = 0$,  
 is classified into 
\begin{eqnarray}
 {\psi_{\bf 16}}_{(++)}&~~~&\mbox{(zero mode)~}=Q,\overline{U},
\overline{E}, \nonumber\\
 {\psi_{\bf 16}}_{(+-)}&~~~&\mbox{(zero mode)}=L,\overline{D}
,\overline{N}, 
\label{5d}
\end{eqnarray}
where the 2nd $\pm$ sign represents the $T_1$ parity.\\

\par
\noindent
\underline{\bf (1). Model 0}\\
Now let us discuss 
 how to generate 
 fermion mass hierarchies and flavor mixings
 in three models.\cite{HKS}\cite{HKS2}
In three models, 
 the 4D brand-localized 
 Higgs fields, ${H_{\bf 16}}$ and ${H_{\overline{\bf 16}}}$, 
 are introduced at $(0,0)$, which are assumed to take 
 vacuum expectation values (VEVs) of $O(10^{16})$ GeV 
 in the directions of $B-L$. 
We also impose the Peccei-Quinn symmetry and its charge on the multiplets:
 all matter multiplets have its charge $1$, $\mathbf{10}$
 representation Higgs multiplets have its charge $-2$,
 and $\mathbf{16}$ and $\overline{\mathbf{16}}$ representation 
 Higgs multiplets have its charge $-1$. 
The superpotential of the Yukawa sector on the brane at $(0,0)$ is given by
\begin{eqnarray}
\label{W_Y}
\label{Yukawa1}
&& W_Y = \left\{
 \frac{y^u_{ij}}{M_*}{\bf H_{10}} {\bf 16}_i {\bf 16}_j
+ \frac{y^d_{ij}}{M_*}{\bf H'_{10}} {\bf 16}_i {\bf 16}_j
\right\}\delta(z)\delta(\bar{z}),
\end{eqnarray}
in which $M_*$ is a ultraviolet cut-off scale of $O(10^{18})$ GeV. 
The index $i,j=1 \sim 3$ shows the generation numbers. 
We consider all components of the Yukawa 
 couplings in Eq.(\ref{Yukawa1}) are 
 of order one.

As for the bulk matter fields, 
 we introduced 5D bulk fields on $\bar{z} = 0$, 
 $({\psi_{\bf 16}}_{(++)}, {\psi_{\bf 16}}_{(-+)}^c)+
  ({\psi_{\bf \overline{16}}}_{(++)}, 
   {\psi_{\bf \overline{16}}}_{(-+)}^c)$, which contains 
 ${\bf 10}+{\bf \overline{10}}$ of $SU(5)$ 
 as the zero modes, which 
 are regarded as  
 the 5th generation fields denoted by 
 $\psi_{{\bf 16}_{5}} + \psi_{\overline{{\bf 16}}_{5}}$.
The 5D bulk fields have the non-chiral structures
 since the 
 5D $N=1$ SUSY corresponds to the 
 4D $N=2$ SUSY. 
We assume that  
 the 5th generation fields 
 interact only with the 2nd generation matter fields. 
In 6D bulk, we introduce 6D bulk vector-like fields, 
 $({\psi_{\bf 16}}_{(++)(++)}, {\psi_{\bf 16}}_{(-+)(++)}^c)+
  ({\psi_{\bf \overline{16}}}_{(++)(-+)}, 
   {\psi_{\bf \overline{16}}}_{(-+)(-+)}^c)$, 
 $({\psi_{\bf 16}}_{(-+)(++)}, {\psi_{\bf 16}}_{(++)(++)}^c)+
  ({\psi_{\bf \overline{16}}}_{(-+)(-+)}, 
   {\psi_{\bf \overline{16}}}_{(++)(-+)}^c)$ and 
 $({\psi_{\bf 16}}'_{(++)(+-)}, {\psi_{\bf 16}}'^c_{(-+)(+-)})+
  ({\psi_{\bf \overline{16}}}'_{(++)(--)}, 
   {\psi_{\bf \overline{16}}}'^c_{(-+)(--)})$, 
 $({\psi_{\bf 16}}'_{(-+)(+-)}, {\psi_{\bf 16}}'^c_{(++)(+-)})+
  ({\psi_{\bf \overline{16}}}'_{(-+)(--)}, 
   {\psi_{\bf \overline{16}}}'^c_{(++)(--)})$, 
 which contains 
 ${\bf 10}+{\bf \overline{10}}$ of $SU(5)$ 
 as the zero modes, which 
 are regarded as  
 the 4th generation fields denoted by 
 $\psi_{{\bf 16}_{4}} + \psi_{\overline{{\bf 16}}_{4}}$ and
 $\psi'_{{\bf 16}_{4}} + \psi'_{\overline{{\bf 16}}_{4}}$,
 respectively. 
We assume that 
 the 4th generation fields 
 interact with only the 1st generation matter fields.

Then, in addition to the superpotential in Eq.(\ref{Yukawa1}),
 the following interactions between the chiral 
 and extra generation fields on the 4D brane, $(0,0)$, 
\begin{eqnarray}
\label{W6}
W_6 &=& H_{16}H_{\overline{16}} \left\{ 
\frac{y_{44}}{M_*^3}\psi_{{\bf 16}_4}\psi_{\overline{{\bf 16}}_4}
+ \frac{y'_{44}}{M_*^3}\psi'_{{\bf 16}_4}\psi'_{\overline{{\bf 16}}_4}
+ \frac{y_{14}}{M_*^{2}}{\bf 16}_1\psi_{\overline{{\bf 16}}_4}
+ \frac{y'_{14}}{M_*^{2}}{\bf 16}_1\psi'_{\overline{{\bf 16}}_4}
 \right.\nonumber\\
&& \left.
+ \frac{y_{55}}{M_*^2}\psi_{{\bf 16}_5}\psi_{\overline{{\bf 16}}_5}
+ \frac{y_{25}}{M_*^{3/2}}{\bf 16}_2\psi_{\overline{{\bf 16}}_5}\right\}
\delta(z)\delta(\bar{z}).
\end{eqnarray}
Where we assume that the vector-like masses which mix the 4th and the
 5th generations are forbidden by the fundamental theory.
Bellow the compactification scale, 
 the interactions in Eq.(\ref{W6}) induce the mass terms for the 
 Kaluza-Klein zero-modes of vector-like matter fields as, 
\begin{eqnarray}
\label{W_4}
W_4 
&\simeq&
{v_N^2 \over M_*}
 \left\{\epsilon_1^{4}\left(Q^{(0)}_4 \overline{Q}^{(0)}_4
+ U'^{(0)}_4 \overline{U}'^{(0)}_4 + E'^{(0)}_4 \overline{E}'^{(0)}_4
\right) + \epsilon_1^{2} \left(Q_1 \overline{Q}^{(0)}_4  
+\overline{U_1} U'^{(0)}_4 + \overline{E_1} E'^{(0)}_4
\right)\right.  \nonumber\\
&&\left.
+ \epsilon_2^{2} \left(Q^{(0)}_5 \overline{Q}^{(0)}_5 
+ U^{(0)}_5 \overline{U}^{(0)}_5 + E^{(0)}_5 \overline{E}^{(0)}_5
\right) 
+ \epsilon_2 \left(Q_2 \overline{Q}^{(0)}_5 
+ \overline{U_2} U^{(0)}_5 + \overline{E_2} E^{(0)}_5
\right)
\right\},\label{W4}
\end{eqnarray}
where $\langle {H_{\bf 16}}\rangle =
 \langle{H_{\overline{\bf 16}}}\rangle \equiv v_N$. 
$\epsilon_i$s are the volume suppression factors which 
 are given by 
\begin{eqnarray} 
 \epsilon_1 = \epsilon_2 \equiv 1/ \sqrt{{\pi R_c M_*}}. 
\label{eps} 
\end{eqnarray}
These volume suppression factors 
 play crucial roles for generating 
 the fermion mass matrices 
 in the low energy\footnote{
The zero modes of the 4th and 5th generation 
 fields have vector-like masses of
 $\epsilon_1^{4}v_N^2/M_*$ and $\epsilon_2^{2}v_N^2/M_*$, respectively.
Since these zero modes form $SU(5)$ multiples, 
 we can expect that the gauge coupling unification is not spoiled.}.
Now we set $1/R_c = O(10^{16})$ GeV, which means 
 $\epsilon_i \simeq\lambda^2 \sim 0.04$,  
 where $\lambda$ is the Cabibbo angle, $\lambda \sim 0.2$.

After integrating out 
 the heavy fields, 
 the model gives the following mass matrices 
 in the up quark sector, the down quark sector, 
 and the charged lepton sector,\cite{HKS}
\begin{equation}
 m_u^l \simeq \left(
\begin{array}{ccc}
 \lambda^8 & \lambda^6 &  \lambda^4  \\
 \lambda^6 & \lambda^4 &  \lambda^2  \\
 \lambda^4 & \lambda^2 &  1
\end{array}
\right)  v, \;\;
 m_d^l \simeq \left(
\begin{array}{ccc}
\lambda^4 & \lambda^4 &  \lambda^4  \\
\lambda^2 & \lambda^2 &  \lambda^2  \\
  1 & 1 & 1 
\end{array}
\right) \overline{v}, \;\;
 m_e^l \simeq \left(
\begin{array}{ccc}
 \lambda^4 & \lambda^2 & 1  \\
 \lambda^4 & \lambda^2 & 1  \\
 \lambda^4 & \lambda^2 & 1  
\end{array}
\right)  \overline{v}, \;\;
\label{mass}
\end{equation}
respectively. 
 $\overline{v}$ and    
 $v$ are the vacuum expectation values of the weak Higgs doublets.  
We write the mass matrices in the basis
 that the left-handed fermions are to the left 
 and the right-handed fermions are to the right. 
We notice that all elements in the mass matrices 
 have $O(1)$ 
 coefficients.  
The fermion mass 
 hierarchies are given by
\begin{eqnarray}
\qquad\quad 
m_t : m_c : m_u  &\simeq & 1 : \lambda^4 : \lambda^8\;, \\  
m_b : m_s : m_d  &\simeq& m_{\tau} : m_{\mu} : m_e 
\simeq 1 : \lambda^2 : \lambda^4\;,
\end{eqnarray}
with the large $\tan \beta$. 
The mass matrix of three light neutrinos $m_\nu^{(l)}$ through
 the see-saw mechanism\cite{seesaw} is given by 
\begin{equation}
\label{26}
 m_\nu^{(l)} \simeq
 {m_\nu^D m_\nu^D{}^T \over M_R} \simeq 
\left(
\begin{array}{ccc}
 1 & 1 & 1 \\
 1 & 1 & 1 \\
 1 & 1 & 1 
\end{array}
\right)  {v^2 \over M_R}.
\end{equation}
$M_R$ is about 
$10^{14}$ GeV induced from the interaction 
\begin{equation}
\label{Yukawa2}
W_{M_N} = \frac{y^{N}_{ij}}{M_*} H_{\overline{16}}H_{\overline{16}} {\bf 
16}_i {\bf 16}_j \delta(z)\delta(\bar{z})
\end{equation}
at $(0,0)$. 
We can obtain the suitable mass scale ($O(10^{-1})$ eV) for
 the atmospheric neutrino oscillation experiments, 
 by taking account of the $SO(10)$ relation, 
 $y_u \simeq y_{\nu}$.

As for the flavor mixings, 
 the CKM\cite{CKM} and the MNS\cite{MNS} matrices are given by 
\begin{equation}
\label{mixing1}
 V_{CKM} \simeq \left(
\begin{array}{ccc}
 1 & \lambda^2 & \lambda^4 \\
 \lambda^2 & 1 & \lambda^2 \\
 \lambda^4 & \lambda^2 & 1 
\end{array}
\right)  , \;\;
 V_{MNS} \simeq \left(
\begin{array}{ccc}
 1 & 1 & 1 \\
 1 & 1 & 1 \\
 1 & 1 & 1 
\end{array}
\right) . 
\end{equation}
which realize the suitable flavor mixings 
 roughly in order of magnitudes. 
They give us a natural explanation 
 why the flavor mixing in the quark sector 
 is small while the flavor mixing in the 
 lepton sector is large.\cite{BB}\tocite{anarchy}
However, they suggest too small Cabibbo angle and 
 too large $V_{e3}$. 
{}For the suitable values of 
 them, we need suitable choice of $O(1)$ coefficients 
 in mass matrices as in Ref.~\citen{BB}. 
Or, if $O(1)$ coefficients are not determined by 
 a specific reason (symmetry) in the fundamental theory, 
 it is meaningful to 
 see the most probable hierarchies and mixing angles 
 by considering random $O(1)$ coefficients.\cite{anarchy}
Anyway, if the fermion mass hierarchies and 
 flavor mixing angles should determined from the fundamental 
 theory 
 in {\it order} (power of $\lambda$) not by tunings of 
 $O(1)$ coefficients, 
 we should modify this scenario. 
We show two examples of the modifications below. \\

\par
\noindent
\underline{\bf (2). Model I}\\
In the first modification, which we call Model I,\cite{HKS2}
 we introduce the 
 additional vector-like 5D bulk matter fields, which are  
 ${\psi_{\bf 16}}_{-} + {\psi_{\overline{{\bf 16}}}}_{-}
 \equiv \psi''_{{\bf 16}_{4}} + 
 \psi''_{{\overline{\bf 16}}_{4}}$.  
They are called as the 4th generation fields and 
 assumed to 
 interact with only the 1st generation. 
Their PQ charge is $1$ as the other matter fields. 
In this case the following terms are added to Eq.(\ref{W6})
\begin{eqnarray}
W_6 &=& H_{16}H_{\overline{16}} \left\{ 
 \frac{y''_{44}}{M_*^2}\psi''_{{\bf 16}_4}\psi''_{\overline{{\bf 16}}_4}
+ \frac{y''_{24}}{M_*^{3/2}}{\bf 16}_1\psi''_{\overline{{\bf 16}}_4}\right\}
\delta(z)\delta(\bar{z}). 
\end{eqnarray}
When $1/R_c = O(10^{16})$ GeV, 
 which means $\epsilon_1 \sim \epsilon_2 \sim \lambda^2$, 
 the fermion mass matrices in the low energy\cite{lop} are given by 
\begin{eqnarray}
 &m_u^l& \simeq \left(
\begin{array}{ccc}
 \lambda^8 & \lambda^6 &  \lambda^4  \\
 \lambda^6 & \lambda^4  &  \lambda^2  \\
 \lambda^4  & \lambda^2  & 1
\end{array}
\right)  v, \;\;
 m_d^l \simeq \left(
\begin{array}{ccc}
 \lambda^6 & \lambda^4 &  \lambda^4  \\
 \lambda^4 & \lambda^2  &  \lambda^2  \\
  \lambda^2 & 1 & 1 
\end{array}
\right) \overline{v}, \;\; \nonumber \\
 &m_e^l& \simeq \left(
\begin{array}{ccc}
 \lambda^6 & \lambda^4 & \lambda^2  \\
 \lambda^4 & \lambda^2  & 1 \\
 \lambda^4 & \lambda^2 & 1 
\end{array}
\right)  \overline{v}, \;\;
 m_\nu^{(l)} \simeq\left(
\begin{array}{ccc}
\lambda^4 & \lambda^2 & \lambda^2 \\
\lambda^2 & 1 & 1 \\
\lambda^2 & 1 & 1
\end{array}
\right)  {v^2 \over M_R},
\end{eqnarray}
after integrating out the heavy vector-like fields. 
They induce the more realistic fermion 
 mass hierarchies as 
\begin{eqnarray}
\label{mass1}
\qquad\quad 
m_t : m_c : m_u  &\simeq & 1 : \lambda^4 : \lambda^8\;, \nonumber\\
m_b : m_s : m_d  &\simeq& m_{\tau} : m_{\mu} : 
m_e \simeq 1 : \lambda^2 : \lambda^6\;,
\end{eqnarray}
with large $\tan \beta$. 
As for the neutrino sector, 
 the rank of $2 \times 2$ sub-matrix in the 2nd and the 
 3rd generations in 
 $m_{\nu}^{(l)}$ should be reduced, and the 
 light eigenvalue of this sub-matrix should 
 be of $O(\lambda^2)$ for the LMA solar neutrino solution. 
This case induce the hierarchical type of 
 neutrino mass, $m_1, m_2 \ll m_3$.\cite{vissani}
Then, 
 the CKM and the MNS matrices become 
\begin{equation}
 V_{CKM} \simeq \left(
\begin{array}{ccc}
 1 & \lambda^2 & \lambda^4 \\
 \lambda^2 & 1 & \lambda^2 \\
 \lambda^4 & \lambda^2 & 1 
\end{array}
\right) , \;\;
 V_{MNS} \simeq \left(
\begin{array}{ccc}
 1/\sqrt{2} & 1/\sqrt{2} & \lambda^2 \\
 1/2 & -1/2 & 1/\sqrt{2} \\
 -1/2 & 1/2 & 1/\sqrt{2} 
\end{array}
\right), 
\end{equation}
where the MNS matrix has the large 1-2 and 2-3 mixings 
 because of the assumption of the rank reduction. 
This case induce small 
 value of $U_{e3}$. 
The CKM matrix has the same structure as in the 
 Model 0. 
Needless to say, the suitable $V_{us}$ can 
 be easily obtained by choosing 
 the O(1) coefficients. \\

\par
\noindent
\underline{\bf (3). Model II}\\
Here let us show the second modification, 
 which we call Model II.\cite{HKS2}
We introduce
 the following bulk matter fields with PQ charge 1 
 in addition to the Model 0:
${\psi_{\bf 16}}_{-+} + {\psi_{\overline{{\bf 16}}}}_{-+} 
\equiv \psi'''_{{\bf 16}_{4}} + 
\psi'''_{{\overline{\bf 16}}_{4}}$ and 
${\psi_{\overline{{\bf 16}}}}_{--} + {\psi_{\overline{{\bf 16}}}}_{--}
 \equiv \psi''''_{{\bf 16}_{4}} + 
\psi''''_{{\overline{\bf 16}}_{4}}$ 
 (we call them the 4th generation fields) 
 which propagate in the 6D bulk and 
 interact with only the 1st generation matter multiplet, 
${\psi_{\bf 16}}_{-} + {\psi_{\overline{{\bf 16}}}}_{-} 
\equiv \psi'_{{\bf 16}_{5}} + 
\psi'_{{\overline{\bf 16}}_{5}}$ 
 (we call them the 5th generation fields)
 which propagate in the 5D bulk ($\bar{z} = 0$) and 
 interact with only the 2nd generation matter multiplet,
${\psi_{\bf 16}}_{-} + {\psi_{\overline{{\bf 16}}}}_{-}  
\equiv \psi_{{\bf 16}_{6}} + 
\psi_{{\overline{\bf 16}}_{6}}$
(we call them the 6th generation fields)
which propagate in the 5D bulk ($\bar{z} = 0$) and 
interact with only the 3rd generation matter multiplet.
In this case, the following terms are added to Eq.(\ref{W6}),
\begin{eqnarray}
W_6 &=& H_{16}H_{\overline{16}} \left\{ 
\frac{y'''_{44}}{M_*^3}\psi'''_{{\bf 16}_4}\psi'''_{\overline{{\bf 16}}_4}
+ \frac{y''''_{44}}{M_*^3}\psi''''_{{\bf 16}_4}\psi''''_{\overline{{\bf 16}}_4}
+ \frac{y'''_{14}}{M_*^{2}}{\bf 16}_1\psi'''_{\overline{{\bf 16}}_4}
+ \frac{y''''_{14}}{M_*^{2}}{\bf 16}_1\psi''''_{\overline{{\bf 16}}_4}
 \right.\nonumber\\
&+& \left.
 \frac{y'_{55}}{M_*^2}\psi'_{{\bf 16}_5}\psi'_{\overline{{\bf 16}}_5}
+ \frac{y'_{25}}{M_*^{3/2}}{\bf 16}_2\psi'_{\overline{{\bf 16}}_5}
+ \frac{y_{66}}{M_*^2}\psi_{{\bf 16}_6}\psi_{\overline{{\bf 16}}_6}
+ \frac{y_{36}}{M_*^{3/2}}{\bf 16}_3\psi_{\overline{{\bf 16}}_6}
\right\}\delta(z)\delta(\bar{z}),~~~~~~ 
\end{eqnarray}
When $1/R_c = O(10^{16})$ GeV, 
 which means $\epsilon_1 \sim \epsilon_2 \sim \lambda^2$, 
 the fermion mass matrices bellow the compactification scale 
 become 
\begin{eqnarray}
 &m_u^l& \simeq \;\left(
\begin{array}{ccc}
 \lambda^8 & \lambda^6 &  \lambda^4  \\
 \lambda^6 & \lambda^4  &  \lambda^2  \\
 \lambda^4  & \lambda^2  & 1
\end{array}
\right)  v, \;\;
 m_d^l \;\;\simeq  \lambda^2 \;\left(
\begin{array}{ccc}
 \lambda^6 & \lambda^4 &  \lambda^4  \\
 \lambda^4 & \lambda^2  &  \lambda^2  \\
  \lambda^2 & 1 & 1 
\end{array}
\right) \overline{v}, \;\; \nonumber \\
 &m_e^l& \simeq  \lambda^2 \left(
\begin{array}{ccc}
 \lambda^6 & \lambda^4 & \lambda^2  \\
 \lambda^4 & \lambda^2  & 1 \\
 \lambda^4 & \lambda^2 & 1 
\end{array}
\;\right)  \overline{v}, \;\;
 m_\nu^{(l)} \simeq \lambda^4 \left(
\begin{array}{ccc}
\lambda^4 & \lambda^2 & \lambda^2 \\
\lambda^2 & 1 & 1 \\
\lambda^2 & 1 & 1
\end{array}
\;\right)  {v^2 \over M_R}.
\end{eqnarray}
The forms of these mass matrices are the same as 
 those of the first case of Model I except for 
 the overall factors. 
Thus the suitable fermion mass hierarchies 
 of the quark and the charged lepton sectors 
 are the same as Eq.(\ref{mass1}). 
The flavor mixing matrices, $V_{CKM}$ and $V_{MNS}$, 
 are also the same as those of the first case 
 of Model I. 
The different between this model and the first case of Model I 
 exists just in the value of 
 $\tan \beta$. 
This model shows the small $\tan \beta$ 
 of $\tan \beta \sim {m_t/m_b \over 1/\lambda^2} \sim 1$. 
The discussion of neutrino mass hierarchy 
 and the flavor mixings are also the same 
 as the first case of Model I.

\section{SUSY breaking and flavor mixings}

Now let us discuss how to induce soft SUSY breaking terms 
 in our model. 
We consider the gaugino and the gauge mediation scenarios
 in Model 0$-$II. 
The soft SUSY breaking terms for the matter fields 
 depend on the two SUSY breaking scenarios and 
 bulk matter configurations
 in the extra dimension. \\

\par
\noindent
\underline{\bf (1). Model 0}\\
As for the gaugino mediation scenario,\cite{KKS}
 SUSY is broken at a spatially different
 place from our living brane, $(0,0)$, in extra dimensions. 
In our 6D theory, there are three fixed points where we can
 put a SUSY breaking field, $S=\theta^2 F$.
Since the gauge multiplets live in the 6D bulk, the gauginos
 receive the SUSY breaking masses through a direct interaction
 with $S$. 
When $S$ is located at the fixed point $(\pi R_z, 0)$, 
 it is possible to write an interaction as,
\begin{eqnarray}
{\cal L} = \frac{1}{M_\ast^3}
 \int dz d\bar{z}\int d^2\theta S {\cal W}_i^\alpha {\cal W}_{i\alpha}
\delta(z-\pi R_z)\delta(\bar{z}).
\end{eqnarray}
This interaction gives rise to the gaugino masses as 
$M_{\widetilde{g}_i}=\epsilon_1^2F/M_\ast\equiv \epsilon_1^2
\overline{m}$.
The cases of $S$ at other fixed points 
 can be considered in the same way. 
Since the chiral matter fields ${\bf 16}_i$ are localized 
 on the 4D brane (0,0), 
 they do not couple directly to $S$. 
Thus, 
 ${\bf 16}_i$ receive SUSY breaking masses 
 only through the renormalization 
 effect of the gaugino masses. 
On the other hand, the 6D bulk matter
 fields can have the direct coupling to $S$ as 
\begin{eqnarray}
\label{eq:6dsoft}
{\cal L} = \frac{1}{M_\ast^4}
 \int dz d\bar{z}\int d^4\theta S^\dagger S 
\left(\psi_{{\bf 16}_4}^\dagger \psi_{{\bf 16}_4}+
\psi'^\dagger_{{\bf 16}_4} \psi'_{{\bf 16}_4}+
\psi^\dagger_{\overline{{\bf 16}}_4}\psi_{\overline{{\bf 16}}_4}+
\psi'^\dagger_{\overline{{\bf 16}}_4}\psi'_{\overline{{\bf 16}}_4}
\right)
\delta(z-\pi R_z)\delta(\bar{z}).~~~~~~~~
\end{eqnarray}
This induces the SUSY breaking mass to the zero-modes
of the 6D matter fields as
\begin{eqnarray}
\label{eq:6dsoft2}
{\cal L}_{\rm soft} = 
 -\epsilon_1^4 \overline{m}^2\left(
\widetilde{Q}^{(0)\dagger}_4\widetilde{Q}^{(0)}_4
+
\widetilde{U'}^{(0)\dagger}_4\widetilde{U'}^{(0)}_4
+
\widetilde{E'}^{(0)\dagger}_4\widetilde{E'}^{(0)}_4
+
\widetilde{\overline{Q}}^{(0)\dagger}_4\widetilde{\overline{Q}}^{(0)}_4
+
\widetilde{\overline{U'}}^{(0)\dagger}_4\widetilde{\overline{U'}}^{(0)}_4
+
\widetilde{\overline{E'}}^{(0)\dagger}_4\widetilde{\overline{E'}}^{(0)}_4
\right)~~~~~~~~~~~
\end{eqnarray}
at the compactification scale. 
The 5D bulk matter fields can also have the direct 
 coupling with $S$ as 
\begin{eqnarray}
{\cal L} = \frac{1}{M_\ast^3}
 \int dz d\bar{z}\int d^4\theta S^\dagger S 
\left(
\psi^\dagger_{{\bf 16}_5} \psi_{{\bf 16}_5}+
\psi^\dagger_{\overline{{\bf 16}}_5}\psi_{\overline{{\bf 16}}_5}
\right)
\delta(z-\pi R_z)\delta(\bar{z}),~~~~
\end{eqnarray}
which induces the SUSY breaking for the zero modes as 
\begin{eqnarray}
\label{eq:5dsoft}
{\cal L}_{\rm soft} = 
 -\epsilon_2^2 \overline{m}^2\left(
\widetilde{Q}^{(0)\dagger}_5\widetilde{Q}^{(0)}_5
+
\widetilde{U}^{(0)\dagger}_5\widetilde{U}^{(0)}_5
+
\widetilde{E}^{(0)\dagger}_5\widetilde{E}^{(0)}_5
+
\widetilde{\overline{Q}}^{(0)\dagger}_5\widetilde{\overline{Q}}^{(0)}_5
+
\widetilde{\overline{U}}^{(0)\dagger}_5\widetilde{\overline{U}}^{(0)}_5
+
\widetilde{\overline{E}}^{(0)\dagger}_5\widetilde{\overline{E}}^{(0)}_5
\right).~~~~~~~
\end{eqnarray}
After integrating out the vector-like heavy fields,
 the soft SUSY breaking mass terms of the 
 light matter fields become 
\begin{equation}
 m^2_{\mathbf {\widetilde 10}} = \left(
\begin{array}{ccc}
  \epsilon_1^4& 0 & 0\\
  0 & \epsilon_2^2 & 0\\
  0 & 0 & 0\\
\end{array}
\right)\overline{m}^2,~~
 m^2_{\mathbf {\widetilde {\overline 5}}} = \left(
\begin{array}{ccc}
  0& 0 & 0\\
  0 & 0 & 0\\
  0 & 0 & 0\\
\end{array}
\right),
\label{35}
\end{equation}
where $\widetilde{10}=
(\widetilde{ Q}, \widetilde{\overline{U}},\widetilde{\overline{E}})$ and
${\widetilde {\overline 5}}=(\widetilde{\overline{D}}, {\widetilde L})$.
They are soft masses 
 around the compactification scale.
Due to quantum corrections from the compactification scale
 to the electroweak scale, additional SUSY breaking masses of order 
 the gaugino masses are added into the diagonal 
 elements of Eq.(\ref{35}). 
These soft SUSY breaking masses, however, are not 
 phenomenologically acceptable. 
That is because, when we take the gaugino masses, 
 $M_{\widetilde{g}_i}= \epsilon_1^2 \overline{m}$, as $O(10^2)$ GeV, 
 Eq.(\ref{35}) suggests 
 $(m_{\widetilde{10}})_{22}$ = $O(10^2)$ TeV. 
Such a heavy soft mass induces 
 the color instability, that is negative mass squared of the 
 stop through the 2-loop renormalization effects.\cite{decouple}
As for other choices of the locations of 
 $S$, at $(0, \pi R_{\bar{z}})$ or $(\pi R_z, \pi R_{\bar{z}})$, 
 through the 
 interaction in Eq.(\ref{eq:6dsoft}) the SUSY breaking masses for
 the light matter fields become 
\begin{equation}
 m^2_{\mathbf {\widetilde 10}} = \left(
\begin{array}{ccc}
  \epsilon_1^4& 0 & 0\\
  0 & 0 & 0\\
  0 & 0 & 0\\
\end{array}
\right)\overline{m}^2,~~
 m^2_{\mathbf {\widetilde {\overline 5}}} = \left(
\begin{array}{ccc}
  0& 0 & 0\\
  0 & 0 & 0\\
  0 & 0 & 0\\
\end{array}
\right).
\label{36}
\end{equation}
The radiative induced 
 masses of order the gaugino masses 
 are added in the diagonal 
 elements in Eq.(\ref{36}). 
Thus, 
 the mass 
 difference between the first and the second 
 generation left-handed down-type squarks is 
 of order the gaugino mass. 
Notice that 
 the experimental constraint from the $K^0\overline{K}^0$ mixing 
 is given by 
\begin{eqnarray}
\sin^2 2\theta_{12} \left(
\frac{(\Delta m^2_{\tilde d})_{12}}{m^2_{\tilde d}}
\right)^2
\left( \frac{10 \mathrm{TeV}}{m_{\tilde d}}\right)^2 \lsim 1,
\end{eqnarray}
where ${m^2_{\tilde d}}=(m^2_{\tilde d_1}m^2_{\tilde d_2})^{1/2}$.  
In Model 0, 
 the flavor mixing between 
 the first and the second generation 
 left-handed down-type squarks is 
 $\sin \theta_{12}\sim \lambda^2$ from Eq.(\ref{mass}), 
 and ${(\Delta m^2_{\tilde d})_{12}}/{m^2_{\tilde d}}\sim 1$. 
Therefore the large FCNC can be avoided 
 when the gaugino mass are $\geq O(1)$ TeV 
 with $S$ at $(0, \pi R_{\bar{z}})$ or $(\pi R_z, \pi R_{\bar{z}})$.

Next 
 we consider the gauge mediation scenario.\cite{DNNS}
We assume that the messenger sector 
 is localized on the 4D brane $(\pi R_z,0)$ and introduce
 $N$ pairs of vector-like messenger fields,
 ${\mathbf 5}^i_M$ and ${\mathbf \overline{5}}^i_M$, 
 which are $5$ and $\overline{5}$ representations
 of the $SU(5)$ group, respectively\footnote{Here 
 we do not specify the dynamical SUSY 
 breaking sector since the mass spectra of 
 the MSSM fields do not depend 
 the detail of them. }.
A $U(1)$ in the bulk can transmit 
 the SUSY breaking effects.\cite{NY}
We consider the following superpotential for 
 the messenger sector.
\begin{equation}
W=\sum_\alpha^N \lambda_\alpha S\, {\mathbf 5}^\alpha_M 
{\mathbf \overline{5}}^\alpha_M
\delta(z-\pi R_z)\delta(\bar{z}),
\end{equation}
where $S=M+\theta^2F_S$ is a spurion
superfield, which represents the SUSY breaking. 
We assign the vanishing PQ charge for ${\mathbf 5}^\alpha_M$, 
${\mathbf \overline{5}}^\alpha_M$, and $S$. 
The gaugino and sfermion masses are 
 induced to the MSSM sector through 
 the SM gauge interactions as
\begin{eqnarray}
\label{messenger}
 M_{\widetilde {g}_a} &\simeq& N \frac{\alpha_a}{4\pi} \frac{F_S}{M},
\\\nonumber 
 (m_{\tilde f}^2)_{ij} &\simeq& 2 N \sum_a
C_a(\tilde {f})
\left(\frac{\alpha_a}{4\pi}\right)^2
 \left|\frac{F_S}{M}\right|^2\delta_{ij},
\end{eqnarray}
where $a(=1-3)$ represents the gauge groups and $C_a(\tilde {f})$
is the quadratic Casimir for the sfermions.
Since the sfermion masses are determined by the gauge quantum
number, the SUSY breaking masses are the same for the vector-like
heavy fields. As a result, the SUSY breaking masses for
the light matter fields are universal around the messenger 
scale. 
It is because 
 the vector-like matter fields in the bulk do not 
 have the coupling of 
 $W=\frac{1}{M_\ast^3}S\,\psi_{{\bf 16}_5}\psi_{\overline{{\bf 16}}_5}
\delta(z-R_z)\delta(\bar{z})$ 
 in the superpotential due to the PQ symmetry 
 in which all matter fields have charge 1. 
The non-universal contribution
 to the sfermion masses 
 is induced from the 
 interaction, 
 $\frac{1}{M_\ast^4}
 \int dz d\bar{z}\int d^4\theta \,S^\dagger S\,
\psi_{{\bf 16}_4}^\dagger \psi_{{\bf 16}_4}
\delta(z-\pi R_z)\delta(\bar{z}) \sim \epsilon_1^4 (\frac{F_S}{M_\ast})^2
\widetilde{Q}^{(0)\dagger}_4\widetilde{Q}^{(0)}_4$.
However, such a non-universal effect is 
$\epsilon_1^4 (\frac{F_S}{M_\ast})^2 \simeq O(10^{-14})\, 
{\mathrm GeV^2}$ with $\sqrt{F_S}=O(10^7)$ GeV, 
 which is negligible 
 compared to the contribution in Eq.(\ref{messenger}).
Thus, the flavor mixing for sfermions 
 are naturally suppressed as in the 
 4D gauge mediation scenario. \\

\par
\noindent
\underline{\bf (2). Model I}\\
In Model I, we consider 
 the gaugino mediation scenario 
 with $S$ being localized on the 4D brane
 $(\pi R_z,0)$ at first. 
The 6D matter fields obtain the SUSY breaking masses 
 as in Eq.(\ref{eq:6dsoft2}). 
For 5D matter fields, 
 $\psi_{{\bf 16}_5}$ and $\psi_{\overline{{\bf 16}}_5}$,
 the SUSY breaking masses are induced as in Eq.(\ref{eq:5dsoft}).
In addition, the interaction between 
 $\psi''_{{\bf 16}_4}$, $\psi''_{\overline{{\bf 16}}_4}$ and 
 $S$, 
\begin{eqnarray}
{\cal L} = \frac{1}{M_\ast^3}
 \int dz d\bar{z}\int d^4\theta S^\dagger S 
\left(
\psi''^\dagger_{{\bf 16}_4} \psi''_{{\bf 16}_4}+
\psi''^\dagger_{\overline{{\bf 16}}_4}\psi''_{\overline{{\bf 16}}_4}
\right)
\delta(z-\pi R_z)\delta(\bar{z}),
\end{eqnarray}
induces the SUSY breaking terms 
\begin{eqnarray}
\label{eq:5dsoft2}
{\cal L}_{\rm soft} = 
 -\epsilon_2^2 \overline{m}^2\left(
\widetilde{L''}^{(0)\dagger}_5\widetilde{L''}^{(0)}_5
+
\widetilde{D''}^{(0)\dagger}_5\widetilde{D''}^{(0)}_5
+
\widetilde{N''}^{(0)\dagger}_5\widetilde{N''}^{(0)}_5
\right.\\\nonumber
\left.+
\widetilde{\overline{L''}}^{(0)\dagger}_5\widetilde{\overline{L''}}^{(0)}_5
+
\widetilde{\overline{D''}}^{(0)\dagger}_5\widetilde{\overline{D''}}^{(0)}_5
+
\widetilde{\overline{N''}}^{(0)\dagger}_5\widetilde{\overline{N''}}^{(0)}_5
\right).
\end{eqnarray}
After integrating out the vector-like heavy fields, 
 the soft SUSY breaking 
 masses for the light matter fields are given by 
\begin{equation}
 m^2_{\mathbf {\widetilde 10}} = \left(
\begin{array}{ccc}
  \epsilon_1^4& 0 & 0\\
  0 & \epsilon_2^2 & 0\\
  0 & 0 & 0\\
\end{array}
\right)\overline{m}^2,~~
 m^2_{\mathbf {\widetilde {\overline 5}}} = \left(
\begin{array}{ccc}
  \epsilon_2^2& 0 & 0\\
  0 & 0 & 0\\
  0 & 0 & 0\\
\end{array}
\right)\overline{m}^2.~~
\label{42}
\end{equation}
They suggest that the above mass spectra 
 suffer from the large FCNC problem
 as in Model 0. 
For the cases where $S$ is localized on the 4D brane 
 $(0, \pi R_{\bar{z}})$ or $(\pi R_z, \pi R_{\bar{z}})$, 
 the FCNC problem can be solved 
 with gaugino mass $\geq O(1)$ TeV, 
 since $\epsilon_2^2$s vanish in Eq.(\ref{42}). 
These are the same situations as 
 in Model 0. 
While the gauge mediation mechanism 
 works well as in Model 0. \\

\par
\noindent
\underline{\bf (3). Model II}\\
As for the gaugino mediation scenario
 with $S$ being located at $(\pi R_z,0)$ 
 in Model II, 
 the 6D matter fields, 
$\psi_{{\bf 16}_4}$, $\psi_{\overline{{\bf 16}}_4}$,
$\psi'_{{\bf 16}_4}$, and $\psi'_{\overline{{\bf 16}}_4}$,
 obtain the SUSY breaking masses 
 as in Eq.(\ref{eq:6dsoft2}).
For the additional 6D fields,
$\psi'''_{{\bf 16}_4}$, $\psi'''_{\overline{{\bf 16}}_4}$,
$\psi''''_{{\bf 16}_4}$, and $\psi''''_{\overline{{\bf 16}}_4}$,
they have a direct coupling to $S$, 
\begin{eqnarray}
{\cal L} = \frac{1}{M_\ast^4}
 \int dz d\bar{z}\int d^4\theta S^\dagger S 
\left(
\psi'''^\dagger_{{\bf 16}_4} \psi'''_{{\bf 16}_4}+
\psi'''^\dagger_{\overline{{\bf 16}}_4}\psi'''_{\overline{{\bf 16}}_4}
+\psi''''^\dagger_{{\bf 16}_4} \psi''''_{{\bf 16}_4}+
\psi''''^\dagger_{\overline{{\bf 16}}_4}\psi''''_{\overline{{\bf 16}}_4}
\right)
\delta(z-\pi R_z)\delta(\bar{z}),~~~~~~~~
\end{eqnarray}
which induces SUSY breaking masses as
\begin{eqnarray}
\label{eq:6dsoft3}
{\cal L}_{\rm soft} = 
 -\epsilon_1^4 \overline{m}^2\left(
\widetilde{L''''}^{(0)\dagger}_4\widetilde{L''''}^{(0)}_4
+
\widetilde{D'''}^{(0)\dagger}_4\widetilde{D'''}^{(0)}_4
+
\widetilde{N'''}^{(0)\dagger}_4\widetilde{N'''}^{(0)}_4
\right.\\\nonumber
\left.
+
\widetilde{\overline{L''''}}^{(0)\dagger}_4\widetilde{\overline{L''''}}^{(0)}_4
+
\widetilde{\overline{D'''}}^{(0)\dagger}_4\widetilde{\overline{D'''}}^{(0)}_4
+
\widetilde{\overline{N'''}}^{(0)\dagger}_4\widetilde{\overline{N'''}}^{(0)}_4
\right).
\end{eqnarray}
On the other hand, 
 the 5D matter fields have the interaction, 
Eq.(\ref{eq:5dsoft}).
\begin{eqnarray}
{\cal L} = \frac{1}{M_\ast^3}
 \int dz d\bar{z}\int d^4\theta S^\dagger S 
\left(
\psi'^\dagger_{{\bf 16}_5} \psi'_{{\bf 16}_5}+
\psi'^\dagger_{\overline{{\bf 16}}_5}\psi'_{\overline{{\bf 16}}_5}+
\psi^\dagger_{{\bf 16}_6} \psi_{{\bf 16}_6}+
\psi^\dagger_{\overline{{\bf 16}}_6}\psi_{\overline{{\bf 16}}_6}
\right)
\delta(z-\pi R_z)\delta(\bar{z}),~~~~~~~
\end{eqnarray}
which induces the SUSY breaking masses,
\begin{eqnarray}
{\cal L}_{\rm soft} &&= 
 -\epsilon_2^2 \overline{m}^2\left(
\widetilde{L'}^{(0)\dagger}_5\widetilde{L'}^{(0)}_5
+
\widetilde{D'}^{(0)\dagger}_5\widetilde{D'}^{(0)}_5
+
\widetilde{N'}^{(0)\dagger}_5\widetilde{N'}^{(0)}_5
+
\widetilde{\overline{L'}}^{(0)\dagger}_5\widetilde{\overline{L'}}^{(0)}_5
+
\widetilde{\overline{D'}}^{(0)\dagger}_5\widetilde{\overline{D'}}^{(0)}_5
+
\widetilde{\overline{N'}}^{(0)\dagger}_5\widetilde{\overline{N'}}^{(0)}_5
\right.\nonumber\\ 
&&\left.
+\widetilde{L}^{(0)\dagger}_6\widetilde{L}^{(0)}_6
+
\widetilde{D}^{(0)\dagger}_6\widetilde{D}^{(0)}_6
+
\widetilde{N}^{(0)\dagger}_6\widetilde{N}^{(0)}_6
+
\widetilde{\overline{L}}^{(0)\dagger}_6\widetilde{\overline{L}}^{(0)}_6
+
\widetilde{\overline{D}}^{(0)\dagger}_6\widetilde{\overline{D}}^{(0)}_6
+
\widetilde{\overline{N}}^{(0)\dagger}_6\widetilde{\overline{N}}^{(0)}_6
\right).
\end{eqnarray}
After integrating out the vector-like heavy fields, the SUSY breaking 
masses for the light matter fields are given by
\begin{equation}
 m^2_{\mathbf {\widetilde 10}} = \left(
\begin{array}{ccc}
  \epsilon_1^4& 0 & 0\\
  0 & \epsilon_2^2 & 0\\
  0 & 0 & 0\\
\end{array}
\right)\overline{m}^2,~~
 m^2_{\mathbf {\widetilde {\overline 5}}} = \left(
\begin{array}{ccc}
 \epsilon_1^4& 0 & 0\\
  0 & \epsilon_2^2 & 0\\
  0 & 0 & \epsilon_2^2\\
\end{array}
\right)\overline{m}^2.~~
\label{47}
\end{equation}
These mass spectra also suffer from the 
 phenomenological problem
 as in Models 0 and I.
As for the cases where $S$ is localized on the 4D brane 
 $(0, \pi R_{\bar{z}})$ or
 $(\pi R_z, \pi R_{\bar{z}})$, 
 the FCNC problem can be avoidable 
 when gaugino mass $\geq O(1)$ TeV
 as in the Models 0 and I, 
 since $\epsilon_2^2$s vanish in Eq.(\ref{47}). 
These are the same situations as 
 in Models 0 and I.  
On the other hand, the gauge mediation in Model II 
 works well as in Models 0 and I.

\section{Summary and Discussion}

In this 
 paper, we have shown three models based on 
 the 6D $N = 1$ SUSY 
 $SO(10)$ GUT where the 5th and 6th dimensional coordinates are 
 compactified on a $T^2/(Z_2 \times Z_2')$ orbifold. 
The gauge and Higgs fields live 
 in 6 dimensions while ordinal chiral matter fields are 
 localized in 4 dimensions. 
We have shown briefly three models which can 
 produce the suitable fermion mass hierarchies and 
 flavor mixings. 
In these models, 
 the three-generation chiral matter
 fields are localized at the 4D wall, and 
 the suitable fermion mass hierarchies and flavor mixings 
 are generated by integrating out
 vector-like heavy generations.

As for the SUSY breaking mechanisms, 
 we have considered 
 the gaugino and the gauge mediation scenarios. 
In the gaugino mediation scenario, 
 the vector-like matter fields in extra dimensions
 can directly couple to the SUSY breaking fields,
 which induces non-universal contributions to
 SUSY breaking masses for the light matter fields.
These non-universal SUSY breaking masses can give rise to
 too large flavor changing neutral currents (FCNCs).  
Thus, the location of the SUSY breaking brane 
 should be determined in order to avoid the large FCNC 
 phenomenological problems in the gaugino mediation 
 scenario. 
The condition of the gaugino mass $\geq O(1)$ TeV 
 is also needed. 
On the other hand, 
 the SUSY breaking 
 masses for the light matter fields are highly 
 degenerated in the gauge mediation scenario, 
 where the FCNCs 
 are naturally suppressed as in the ordinal 
 4D gauge mediation models.

Finally we comment on other SUSY breaking 
 scenarios. 
The gravity mediation scenario gives rise 
 the non-universal corrections to the 
 soft SUSY masses in general. 
It is because the SUSY breaking effects 
 are mediated by ``Yukawa'' interactions not 
 by gauge interactions. 
The ``Yukawa'' interactions 
 among the bulk fields always receive 
 the volume suppressions, which 
 violate the degeneracy of the soft SUSY breaking masses. 
The Scherk-Schwarz SUSY breaking\cite{SS}\cite{SS2} 
 might also give the non-negligible effects 
 of breaking degeneracy, since 
 the first and the second generation fields
 are mainly composed by the bulk fields in 
 our three models.


\section*{Acknowledgment}
We would like to thank T. Kugo and 
 Y. Nomura for
 helpful discussions. 
This work is supported in
 part by the Grant-in-Aid for Science
 Research, Ministry of Education, Culture, Sports, Science and 
Technology, of Japan (No. 14039207, No. 14046208, No. 14740164).


\end{document}